\documentclass[10pt,superscriptaddress,twocolumn,amsmath,amssymb,aps,pra,showpacs]{revtex4-1}
\usepackage{mathrsfs}
\usepackage{graphicx}
\usepackage{dcolumn}
\usepackage{bm}
\usepackage[title,titletoc,header]{appendix}

\usepackage{amssymb}
\usepackage{amsmath}
\usepackage{mathrsfs}
\usepackage{amsmath}
\usepackage{subfigure}
\usepackage{float}
\usepackage[title,titletoc,header]{appendix}
\usepackage{graphicx}
\usepackage{color}

\newcommand{\be}{\begin{equation}}
\newcommand{\ee}{\end{equation}}
\newcommand{\bea}{\begin{eqnarray}}
\newcommand{\eea}{\end{eqnarray}}

\begin{document}
\title{Strongly Interacting $p$-wave Fermi Gas in Two-Dimensions: Universal Relations and Breathing Mode}
\author{Yi-Cai Zhang}
\affiliation{Department of Physics and Center of Theoretical and
Computational Physics, The University of Hong Kong, Hong Kong,
China}

\author{Shizhong Zhang}
\affiliation{Department of Physics and Center of Theoretical and
Computational Physics, The University of Hong Kong, Hong Kong,
China}

\date{\today}
\begin{abstract}
The contact is an important concept that characterizes the universal properties of a strongly interacting quantum gas. It appears in both thermodynamic (energy, pressure, {\em etc.}) and dynamic quantities (radio-frequency and Bragg spectroscopies, {\em etc.}) of the system. Very recently, the concept of contact has been extended to higher partial waves, in particular, the $p$-wave contacts have been experimentally probed in recent experiment. So far discussions on $p$-wave contacts have been limited to three-dimensions. In this paper, we generalize the $p$-wave contacts to two-dimensions and derive a series of universal relations, including the adiabatic relations, high momentum distribution, virial theorem and pressure relation.  At high temperature and low density limit, we calculated the $p$-wave contacts explicitly using virial expansion. A formula which directly connects the shift of the breathing mode frequency and the $p$-wave contacts are given in a harmonically trapped system. Finally, we also derive the relationships between interaction parameters in three and two dimensional Fermi gas and discuss possible experimental realization of two dimensional Fermi gas with $p$-wave interactions.
\end{abstract}

\maketitle

\section{ Introduction}
\label{intro}
In cold atomic gas, the resonance $p$-wave interaction was realized experimentally quite some time ago in both $^{40}$K and $^6$Li~\cite{Regal2003,Zhang2004,Ticknor,Gunter2005,Schunck2005,Gaebler2007,Inada2008,Fuchs2008,Nakasuji2013,Waseem2016}. It was observed that the system in general suffers significant loss close to resonance, preventing the systematic study of the many-body system in equilibrium and in particular, the achievement of superfluid regime. Recently, radio-frequency spectroscopic measurement in $^{40}$K shows that the normal state of $p$-wave Fermi gas close to resonance can achieve a quasi-equilibrium state in which the equilibration between scattering fermions and the shallow $p$-wave dimers within the $p$-wave centrifugal barrier, thus establishing a strongly interacting $p$-wave system \cite{Luciuk}. The extracted free energy reduction close to resonance is of order of the Fermi energy. Furthermore, it was shown that just as in the $s$-wave case \cite{Tan2008,Braaten,shizhong2009,Werner2009}, the $p$-wave Fermi gas also show universal relations, except that now both the $p$-wave scattering volume $v$ and the effective range $R$ are relevant. Apart from this, there now appear several molecular states due to the different orientation of the angular momentum quantization of the molecule. As a result, the contact parameters has to be generalized which leads to a set of universal relations that has been discussed in detail in recent literature\cite{Yoshida,Yu2015,He,Yoshida2016,Peng2016,Cui2016,Zhang2016}. Explicit calculations of the $p$-wave contact within the Nozi\'{e}res and Schmit-Rink formula has been carried out and finds general good agreement with the experimental findings \cite{Yaojuan}.

So far, $p$-wave contacts has been defined and discussed mostly in the three-dimensional case. In two-dimensions, the derivation of these universal relations proceed essentially in the same manner as in three-dimension, except that while in three-dimensions, one is dealing with a power-law divergence; in two-dimension, one has to deal with the logarithmic divergence. This requires a slight generalization especially in dealing with the $p$-wave effective range. Another interesting aspect of two-dimensional system is the apparent scale invariance in the $s$-wave delta-function interaction \cite{Pitaevskii}, which would predict a breathing mode frequency exactly at twice the trap frequency. Here we investigate the similar problem in 2D $p$-wave resonance and derive the equation of motion for the breathing mode and show that the $p$-wave contact is also implicated in the equation of motion. In particular, we show that at resonance, the contact parameter related to the effective range breaks the scale invariance and determines the breathing mode frequency shift.

In this paper, we extend the concept of $p$-wave contacts to two dimensions. In the Sec. \textbf{\ref{tbs}}, we review some basic facts about low energy scattering and in particular, the scattering amplitude for $p$-wave interaction and its associated weakly bound state. In Section \textbf{\ref{definition}}, we define the $p$-wave contacts in two-dimensions and derive the adiabatic theorem. Universal relations for the tail of momentum distribution, the virial theorem and pressure relation are given in Section \textbf{\ref{universal}}. In Section \textbf{\ref{ec}}, we give explicit calculation of the two contacts in two special cases: two-body bound state and high temperature. We apply the theory to the trapped case in Section \textbf{\textbf{\ref{bm}}} and derive the frequency shift of the breathing mode in terms of the $p$-wave contact. General expression for the frequency shift is obtained and its explicit evaluation is given at high temperature. We give a summary in Section \textbf{\ref{summary}}. There are two appendices which discuss the detailed derivations of the frequency shift of the breathing mode and virial theorem in a trap and the relation between effective $p$-wave scattering parameters in two dimensions and that of three dimension.

\section{Two-body scattering and bound states}
\label{tbs}
For spinless Fermi gas, the $s$-wave interaction is totally suppressed due to the Pauli principle. So at low energy, it is the $p$-wave scattering channel that dominates. For a short-range potential, as is usually the case in cold atom experiment, the effective range expansion for $p$-wave in two dimensions is~\cite{Randeria,Hammer2009,Hammer2010,Rakityansky}
\begin{equation}
k^2[\cot \delta_1-\frac{2}{\pi}\ln(\rho k)]=-\frac{1}{a}+\frac{1}{2}r_1 k^2+O(k^4),
\end{equation}
where $\delta_1$ is the $p$-wave phase shift and $r_1$ is a dimensionless parameter;  $a$ and $\rho$ are scattering area and effective range, respectively.
The above equation can be rewritten as
\begin{equation}
 k^2 \cot \delta_1=-\frac{1}{a}+\frac{2k^2}{\pi}\ln(R k) +O(k^4).\label{ere}
\end{equation}
Hereafter we will refer to $R\equiv\rho \exp({\pi r_1}/{4})$ as the $p$-wave effective range.

The Schr\"{o}dinger equation for the relative motion of the two scattering fermions in two-dimension is given by (setting $\hbar=1$ and mass $M=1$)
 \begin{equation}
 \left[\frac{1}{r}\frac{\partial}{\partial r}\left(r\frac{\partial}{\partial r}\right)+\frac{1}{r^2}\frac{\partial^2}{\partial\theta^2}+k^2-U(r)\right]\psi(r,\theta)=0.
\end{equation}
The radial and angular parts of the wave function $\psi(r,\theta)$ can be separated for a central potential $U(r)$, $\psi(r,\theta) \equiv R_k(r)T(\theta)$, where $T(\theta)$ satisfies the following equation
\begin{equation}
\frac{d^2T}{d\theta^2}+m^2T=0,
\end{equation}
and its solution is given by
\begin{equation}
T(\theta)=\frac{1}{\sqrt{2\pi}}e^{\pm im\theta},\notag\\
\end{equation}
where $m\in Z$ is quantum number of angular part and $m=\pm1$ for $p$-wave scattering. The radial part of the wave function satisfies
\begin{equation}
\frac{1}{r}\frac{d}{d r}\left(r\frac{d R(r)}{d r}\right)+\left[k^2-U(r)-\frac{m^2}{r^2}\right]R_k(r)=0.
\end{equation}
Let the range of the potential $U(r)$ be $r_0$. Then for $r>r_0$, the radial wave function $R_k(r)$ can be written as a linear combination of two linearly independent Bessel function
\begin{equation}
R_k(r)= \frac{\pi k}{2}\left[\cot\delta_m(k) J_m(kr)-N_m(kr)\right].\label{radial}
\end{equation}
Here $J_m$ and $N_m$ are the Bessel functions of first and second kinds.
When $r\gg1/k$, using the asymptotic expressions for Bessel functions, the wave function $R_k(r)$ becomes
\begin{equation}
R_k(r)\!\!\propto \!\!\frac{1}{2} \sqrt{\frac{2}{\pi kr}}\left[Se^{i(kr-\frac{m\pi}{2}-\frac{\pi}{4})}+e^{-i(kr-\frac{m\pi}{2}-\frac{\pi}{4})}\right],
\end{equation}
where we have defined the S-matrix in terms of phase shift $\delta_m$,
\begin{equation}
S\equiv\frac{1+i\tan\delta_m}{1-i\tan\delta_m}=e^{2i\delta_m}.
\end{equation}

The scattering wave function can be written as
\begin{equation}
\psi=e^{ikx}+\frac{f}{\sqrt{r}}e^{ikr}=e^{ikr\cos\theta}+\frac{f}{\sqrt{r}}e^{ikr},
\end{equation}
where $f\equiv\sum \hat{f}_m e^{im\theta}/\sqrt{2\pi}$ is scattering amplitude in two dimension.
Taking the $m$-th angular component of the scattering wave function,
\begin{equation}
\psi_m= i^{ m}J_m(kr)e^{i m\theta}+\frac{\hat{f}_m}{\sqrt{r}} e^{i m\theta}e^{ikr},
\end{equation}
and compare with $R_k(r)$ when $kr\gg1$, we get the scattering amplitude of $m$-th wave as
\begin{equation}
\hat{f}_m=e^{i\pi/4}\sqrt{\frac{4}{k}}\frac{1}{\cot\delta_m(k)-i}.
\end{equation}
For $p$-wave scattering with $m=\pm1$ in two-dimensions, using the effective range expansion eqn.(\ref{ere}), we obtain
\begin{equation}
\hat{f}_{\pm1}=e^{i\pi/4}\sqrt{\frac{4}{ k}}\frac{k^2}{-\frac{1}{a}+\frac{2k^2}{\pi}\ln(R k)-ik^2}.\label{sa1}
\end{equation}
The total scattering cross section (taking account of degeneracy of $m=\pm1$)
\begin{equation}
\sigma=\int^{2\pi}_0\left[\left|\hat{f}_{-1}\right|^2+\left|\hat{f}_1\right|^2\right]d\theta=\frac{16\pi \sin^2(\delta_1)}{k}.
\end{equation}

\begin{figure}
\begin{center}
\includegraphics[width=\columnwidth]{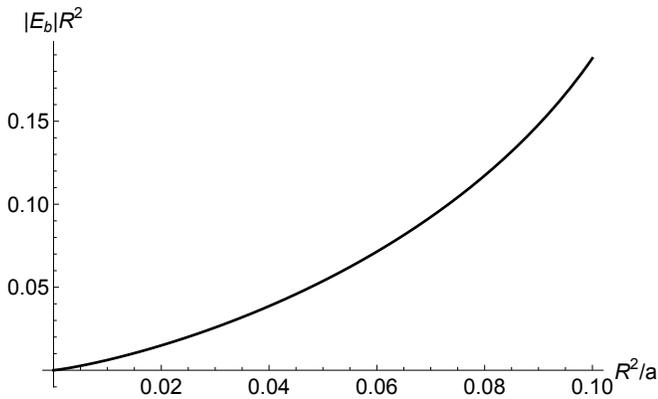}
\end{center}
\caption{(Color online). The variations of binding energy of bound state  $|E_b|$ with $1/a$ (fixed R). Near the resonance $1/a\rightarrow0_+$, the bound state energy approaches zero. }
\label{bsenergy}
\end{figure}

As we shall show later, the effective range $R$ is always positive, and as a result, it is possible that a shallow two-body bound state emerges when $a>0$. The binding energy of the shallow two-body bound state must be much smaller than $1/r_0^2$ in order to be consistent with effective range expansion. Let $k=i\kappa$, one finds that the imaginary part in the denominator of eqn.(\ref{sa1}) vanishes identically, and the real part is zero when ($a>0$)
\begin{equation}
x^2\ln(x)=-\frac{\pi}{2}\frac{R^{2}}{a},\label{bs}
\end{equation}
where $x=R \kappa$ and the bound state energy is given by $E_b=-\kappa^2$. The above equation can be solved using  Lambert W-function, which is $x=\exp[{\frac{1}{2}W_{-1}(-{\pi R^{2}}/{a})}]$. In the limit when $1/a\to 0_+$, one finds that $E_b\simeq {\pi}/({a \ln(\pi R^{2}/a)})$ and $E_b$ tends to $0_-$. The other solution of eqn.(\ref{bs}) is of order or larger than $1/r_0^2$ and has to be discarded. The energy of the shallow bound state is shown in Fig.\ref{bsenergy}.

At low energy when $k\ll1/r_0$, the radial wave function $R_{k}$ can be expanded in terms of powers of $k^2$, {\it e.g.}, $R_k(r)=u_0(r)+k^2u_1(r)+O(k^4)$ and each satisfy
\begin{align}
\left[-\frac{1}{r}\frac{d}{d r}\left(r\frac{d }{d r}\right)+U(r)+\frac{1}{r^2}\right]u_0(r) &=0,\\
\left[-\frac{1}{r}\frac{d}{d r}\left(r\frac{d }{d r}\right)+U(r)+\frac{1}{r^2}\right]u_1(r) &=k^2u_0(r).
\end{align}
For $r\gg r_0$, the radial wave function $R_k(r)$ is given in eqn.(\ref{radial}). Using effective range expansion for scattering phase shift in eqn.(\ref{ere}) and considering the regime where $r_0\ll r\ll 1/k$, the radial wave function, and hence $u_{0,1}(r)$ can be written as
\begin{align}
u_0(r) &=-\frac{1}{r}+\frac{\pi r}{4a}\label{u0},\\
u_{1}(r) &=\frac{r}{2}\ln\frac{r}{\tilde{R}}-\frac{\pi r^3}{32a}\label{u1},
\end{align}
where $\tilde{R}=2\exp({1/2-\gamma_E})R$ and $\gamma_E\simeq0.577$ is the Euler-Mascheroni constant. At the same time, the wave function near the origin {should be} regular which satisfies $R_k(r=0)=0$ and $(r{\partial R_k}/{\partial r})|_{r=0}=0$ {for not very singular inter-atomic potential}.

A Bethe's integral formula for three dimension s-wave effective range \cite{Bethe} has been generalized to arbitrary partial waves \cite{Madsen} and arbitrary dimensions \cite{Hammer2009}. For our two dimensional p-wave case, it takes following form
\begin{align}
\int^{r_c}_{0} ru^{2}_{0}(r)dr=\ln\frac{e^{\gamma_E}r_c/2}{R}-\frac{\pi r^{2}_{c}}{4a}+\frac{ \pi^2 r^{4}_{c}}{64a^{2}},\label{Bethe}
\end{align}
where $r_c\gg r_0$ a cut-off length scale. Note that the normalization of the wave function is slightly different (up to a factor $\sqrt{{\pi}/{2r}}$) from that in reference \cite{Hammer2009}. Because left-hand side of eqn.(\ref{Bethe}) is positive definite, the effective range is bounded by $0<R<e^{\gamma_E}r_0/2\simeq0.89r_0$ close to resonance ($a\to \infty$). Note that in the following,  we will always refer to ``resonance" as $1/a=0$ just as in the three dimensional case where $s$-wave scattering length $1/a_s\rightarrow 0$. The Bethe's formula for effective range [eqn.(\ref{Bethe})] will play an important role in the derivation of breathing mode frequency shift in Sec.\ref{bm}.

Before we end this section, we would like to derive two relations that relate the change of scattering parameters to  the variation of inter-atomic potential $U(r)\to U(r)+\delta U(r)$.  As in the case of three-dimensional $p$-wave scattering~\cite{Yu2015}, two important relations related to the scattering parameters ($a$ and $R$) can be obtained
\begin{align}
\delta(1/a) &=-\frac{2}{\pi}\int_0^{\infty}r\delta (U)u^{2}_{0}(r)dr,\label{inva}\\
\delta(\ln R) &=2\int_0^{\infty}r\delta (U)u_0(r)u_1(r)dr\label{invR}.
\end{align}

\section{Defining the $p$-wave Contacts}
\label{definition}
To define the $p$-wave contacts in two dimension, we follow the route of Ref.~\cite{Yu2015}. Let us consider the two body density matrix
\begin{align}
\rho_2(\vec{r}_1,\vec{r}_2;\vec{r}'_2,\vec{r}'_1)&\equiv\langle\psi^\dag(\vec{r}_1)\psi^\dag(\vec{r}_2)\psi(\vec{r}'_2)\psi(\vec{r}'_1)\rangle\notag\\
&=\rho^*_2(\vec{r}'_1,\vec{r}'_2;\vec{r}_1,\vec{r}_2)
\end{align}
written in terms of second quantized field operators $\psi$. $\rho_2$ is a Hermitian matrix, so it can be diagonalized
\begin{align}
\rho_2(\vec{r}_1,\vec{r}_2;\vec{r}'_2,\vec{r}'_1)=\sum_i n_i\phi^{*}_{i}(\vec{r}_1,\vec{r}_2)\phi_{i}(\vec{r}'_1,\vec{r}'_2).
\end{align}
Here $n_i$ and $\phi_i$ are its eigenvalues and eigenfunctions. For spinless Fermi gas, the eigenfunction $\phi_i$ is odd under exchanges of two fermions $\phi_i(\vec{r}_1,\vec{r}_2)=-\phi_i(\vec{r}_2,\vec{r}_1)$.
When two particles come very close to each other, the eigenfunction can be expanded using two-body wave function (using translational invariance)
\begin{align}
\phi_{\vec{P},m=\pm1}(\vec{r}_1,\vec{r}_2)=\frac{e^{i\vec{P}\cdot\vec{R}}e^{im\theta}}{\sqrt{2\pi V}}\sum_{k} a_{\vec{P},m,k}R_k(r).
\end{align}
Here $\vec{P}=\vec{p}_1+\vec{p}_2$ and $\vec{R}=(\vec{r}_1+\vec{r}_2)/2$ are center of mass momentum and position. $\theta$ describes the angle of relative co-ordinates $\vec{r}=\vec{r}_1-\vec{r}_2$. $a_{\vec{P},m,k}$ is the expansion coefficients introduced such that (1) $\int d\vec{r}_1d\vec{r}_2|\phi(\vec{r}_1,\vec{r}_2)|=1$ is normalized and (2) that the radial wave function $R_{k}(r)$ in the asymptotic regime $r_0\ll r\ll 1/k$ is given by eqns.(\ref{u0},\ref{u1}). As a result, we have
\begin{align}\nonumber
\rho_2(\vec{r}_1,\vec{r}_2;\vec{r}'_2,\vec{r}'_1)=&\sum_{\vec{P},m}n_{\vec{P},m}\frac{e^{i\vec{P}\cdot(\vec{R}-\vec{R}')}e^{im(\theta-\theta')}}{2\pi V}\\
\times&\sum_{k,k'}a^*_{\vec{P},m,k}a_{\vec{P},m,k'}R_k(r)R_{k'}(r').
\end{align}
Here, $r'$ and $\theta'$ are defined accordingly. As in Ref.~\cite{Yu2015}, we have included possible bound state in the sum over $k$.The interaction energy can be expressed as (short-ranged potential $U(r)$)
\begin{align}
\langle U\rangle &=\frac{1}{2}\int d^2r_1d^2r_2 U(r_1-r_2)\rho_2(r_1,r_2;r_1,r_2),\\\nonumber
&=C_{a}\int dr r U(r)u^{2}_{0}(r)+2C_{R}\int dr r U(r)u_0(r)u_1(r),
\end{align}
where we have defined two $p$-wave contacts $C_{a}$ and $C_{R}$ as
\begin{align}
 C_{a} &\equiv\frac{1}{2}\sum_{P,m,k'k} n_{P,m}a^{*}_{P,m,k'}a_{P,m,k},\\
 C_{R} &\equiv\frac{1}{4}\sum_{P,m,k'k} n_{P,m}a^{*}_{P,m,k'}a_{P,m,k}(k'^2+k^2).
\end{align}
By definition, $C_{a}\geq 0$, while $C_{R}$ has no definite signs. The variations of free energy $F$ as inter-atomic potential are varied by $\delta U$ is given by
\begin{align}
&\delta F =\delta\langle U\rangle,\\\nonumber
&=C_{a}\int dr r \delta Uu^{2}_{0}(r)+2C_{R}\int dr r \delta Uu_0(r)u_1(r).
\end{align}
Using eqns.(\ref{inva}, \ref{invR}), we obtain two adiabatic relations
\begin{align}
\frac{\partial F}{\partial a^{-1}}&=-\frac{\pi}{2}C_{a},\\
\frac{\partial F}{\partial\ln R} &= C_{R}.
\end{align}
Similar to the cases of $s$- and $p$-wave scattering in three dimensions, the above two equations give the important relationships between the two contacts ($C_a, C_R$) and the thermodynamics of the system. Here we should mention that for $p$-wave resonant Fermi gas in two dimensions, the effective potential of resulting from three body correlation could support the so-called super Efimov states, for which the energy for three-body has double exponentials scaling law~\cite{Nishida2013,Volosniev2014,Gao2015}. Very recently, the three-body contact arising from the super Efimov states is discussed in Ref.~\cite{PengfeiZhang}. However, in the present paper, we shall focus on contacts arising from two-body correlations.


\section{Universal relations}
\label{universal}
The utility of contacts lies in that it relates various physical observables and provides a crucial consistency check on the theory and experiments.

\subsection{Tails of momentum distribution}
The derivation of momentum distribution follows that of Ref.~\cite{Yu2015}. The momentum
distribution $n_{\vec{q}}$ is related to single-particle density matrix $\rho_1(r,r')$
\begin{align}
\rho_1(\vec r,\vec r')\equiv\langle \psi^\dag(\vec r)\psi(\vec r')\rangle=\frac{1}{V}\sum_{\vec q} e^{i\vec{q}\cdot \vec{\rho}}n_{\vec q},
\end{align}
where $\vec\rho=\vec r'-\vec r$ and we have used the translational invariance of the system. In terms of many-body wave function $\Psi(\vec{r}_1,\vec{r}_2,\cdots,\vec{r}_N)$, the single-particle density matrix can be written as
\begin{align}
\rho_1(\vec{r}_1,\vec{r}'_1)=N\int d\vec{r}_2\cdots d\vec{r}_N\Psi^*(\vec{r}_1,\vec{r}_2,\cdots,\vec{r}_N)\\\notag\times\Psi(\vec{r}'_1,\vec{r}_2,\cdots,\vec{r}_N).
\end{align}
As a result,
\begin{align}
n_{\vec q} &=\frac{1}{V}\int d^2\vec{r}d\vec{r}' \rho_1(\vec r,\vec r') e^{-i\vec{q}\cdot(\vec{r}-\vec{r}')},\\
&=\frac{N}{V}\int d^2\vec{r}d\vec{r}'d\vec{r}_2\cdots d\vec{r}_N e^{-i\vec{q}\cdot(\vec{r}-\vec{r}')}\Psi^*(\vec{r},\vec{r}_2,\cdots,\vec{r}_N)\notag\\
~~&\times\Psi(\vec{r}',\vec{r}_2,\cdots,\vec{r}_N).
\end{align}
Now, we need $\vec{r}$ to be close to $\vec{r}'$ in order to extract the high momentum distribution. This requires that one of the co-ordinates $\vec{r}_2\cdots\vec{r}_N$ to be close to both $\vec{r}$ and $\vec{r}'$ which gives the singular contribution to one-body density matrix. Note that in total there are $N-1$ possibilities, and we finally obtain $n_{\vec{q}}$ in terms of the singular part of the two-body density matrix
\begin{align}
n_{\vec q} &=\!\! \frac{1}{V^2}\!\!\sum_{\vec Pmk'k}\!\!n_{\vec P,m}a^{*}_{\vec P,m,k'}a_{\vec P,m,k}\!\!\int\!\! d^2 \vec{r}d^2\vec{r}'\!e^{i(\vec{P}/2-\vec{q})\cdot(\vec r-\vec r')} \notag\\
&\left\{\int d^2\vec r_1 \frac{e^{im\theta}}{|\vec r-\vec r_1|}\frac{e^{-im\theta'}}{|\vec r'-\vec r_1|}-(k'^{2}+k^2)\int d^2\vec r_1 \right.\notag\\
&\left.\frac{e^{i\theta}|\vec r-\vec r_1|}{2}\ln\frac{|\vec r-\vec r_1|}{\tilde{R}}\frac{e^{-i\theta'}}{|\vec r'-\vec r_1|}\right\},\\
&=\frac{(2\pi)^2}{V}\sum_{\vec Pmk'k}n_{\vec P,m}a^{*}_{\vec P,m,k'}a_{\vec P,m,k}\notag\\
&\times\left\{\frac{1}{q^2}+\frac{(k'^{2}+k^2)+\vec{P}\cdot\vec{q}-P^2/4}{q^4}\right\}.
\end{align}
That is
\begin{equation}
n_{\vec{q}}\sim\frac{4\pi}{V}\left\{\frac{C_{a}}{q^2}+\frac{2C_{R}-C_P}{q^4}\right\},
\end{equation}
where
\begin{equation}
C_P\equiv\frac{\pi}{4}\sum_{P,m,k'k}n_{P,m}a^{*}_{P,m,k'}a_{P,m,k}P^2,
\end{equation}
which arises from the center of mass motion of the pairs. In the above derivation, we have assumed that the system respects inversion symmetry so that the linear term in $\vec P$ vanishes.

\subsection{Pressure relation and virial theorem}
To derive the pressure relation and the virial theorem, it is enough to invoke the dimensional analysis. The free energy $F$ of the system should be of the form $F= N \epsilon_F f(T/T_F,k_{F}^{2}a,k_FR)$, where $f$ is a dimensionless function, $\epsilon_F=T_F\equiv k^{2}_F/2=2\pi n$ is Fermi energy ($k_B=1$) and, $k_F$ is Fermi momentum and $n$ is particle density. From the thermodynamic relation,
\begin{align}
p &= -\frac{\partial F}{\partial V}=-\frac{\partial F}{\partial k_{\rm F}}\frac{\partial  k_{\rm F}}{\partial V},\\
&=\frac{E}{V}+\frac{1}{V}\frac{\pi C_{a}}{2a}+\frac{C_{R}}{2V}.\label{pressure}
\end{align}
In the last line we have used the adiabatic theorems. Similarly, one can extend the virial theorem in a harmonic trap $V_{\rm trap}(r)=\frac{1}{2}M\omega^2 r^2$, then the total energy can be written as
\begin{equation}
E\equiv E(N,\omega,a,R)=\hbar \omega \varepsilon(N,a/\ell^{2},R/\ell),
\end{equation}
where $\varepsilon$ is a dimensionless function and that $\ell=\sqrt{\hbar/M\omega}$ is the oscillator length. Taking the derivative with respect to the oscillator frequency $\omega$, we have
\begin{equation}
\frac{\partial E}{\partial \omega}=\frac{2\left\langle V_{\rm trap}\right\rangle}{\omega}.
\end{equation}
Carrying out similar manipulations as in the uniform case, we find
\be
2\langle V_{\rm trap}\rangle=E+\frac{\pi}{2}\frac{C_{a}}{a}+\frac{C_{R}}{2}.\label{virial}
\ee
In both eqns.(\ref{pressure}, \ref{virial}), when $a\to \pm \infty$ (at resonance), the second contact $C_R$ breaks the scaling invariance of the two-dimensional $p$-wave Fermi gas at resonance.

\section{Explicit Calculations}
\label{ec}
To gain more insight into the behavior of $p$-wave contacts defined by the adiabatic theorem, we discuss two explicit calculations of them in the following.

\subsection{Two-body bound state}
Let us choose a cutoff scale $r_c$ as before, and write down the wave function for $r\geq r_c$ ($r_c\gg r_0$), in the center of mass coordinates as
\begin{align}
\psi^>(r)=\alpha K_1(\kappa r),
\end{align}
where $\alpha$ is normalization constant and $K_1(r)$ is modified Bessel function of second kind. The bound state energy $E_b=-\kappa^2$. For $0\leq r\leq r_c$, the wave function is
\begin{align}
\psi^<(r)=\alpha\frac{K_1(\kappa r_c)}{u_0(r_c)}u_0(r).
\end{align}
The normalization constant $\alpha$ is determined by, near resonance $1/a\sim 0$
\begin{align}
\frac{1}{\alpha^2}&=2\pi\left\{\int_{0}^{r_c}dr r|\psi^<(r)|^2+\int_{r_c}^{\infty}dr r|\psi^>(r)|^2\right\},\notag\\
&=\frac{2\pi}{\kappa^2 r^{2}_{c}}\frac{1}{(-\frac{1}{r_c}+\frac{\pi r_c}{4a})^2}\left[\ln\frac{e^{\gamma_E}r_c/2}{R}-\frac{\pi r^{2}_{c}}{4a}+\frac{ \pi^2 r^{4}_{c}}{64a^{2}}\right]\notag\\
&+\frac{2\pi}{\kappa^2}[-1/2-\gamma_E+\ln2-\ln(\kappa r_c)+O((\kappa r_c)^2)],\notag\\
&\simeq-\frac{2\pi [\ln(\kappa R)+1/2]}{\kappa^2},
\end{align}
here we have used the Bethe's integral formula. Expanding the wave function {when $\kappa r\ll1$}
\begin{align}
|\psi^>(r)|^2 &=\alpha^2|K_1(\kappa r)|^2,\\
&\simeq\alpha^2\left(\frac{1}{\kappa^2 r^2}+\ln r+...\right),\\
&=-\frac{\kappa^2}{2\pi[\ln(\kappa R)+1/2]}\left(\frac{1}{\kappa^2 r^2}+\ln r+...\right).
\end{align}
at low energy $\kappa R\ll1$. We can now read off the contacts from the above equation
\begin{align}
 C_{a} &=\frac{-1}{ln(\kappa R)+1/2}>0, \\
 C_{R} &=\frac{\kappa^2}{\ln(\kappa R)+1/2}<0.
\end{align}
Using the equation for the bound state energy eqn. (\ref{bs}), one can verify the adiabatic theorems, namely that $\frac{\partial E_b}{\partial a^{-1}}=-\frac{\pi}{2}C_{a}$ and $\frac{\partial E_b}{\partial \ln R}=C_{R}$.

\subsection{Virial expansion}
\begin{figure}
\begin{center}
\includegraphics[width=\columnwidth]{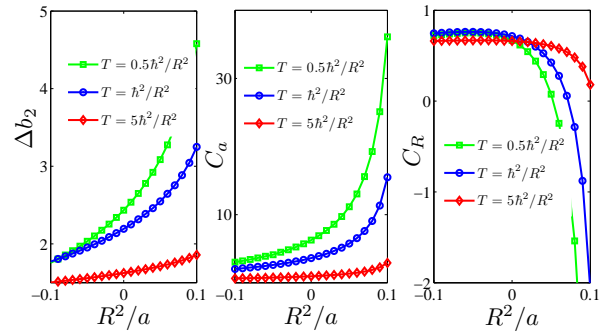}
\end{center}
\caption{(Color online). The variations of contacts $C_{a}$ and $C_{R}$ with the interaction parameter $1/a$ for different temperatures. Here we use the effective range $R$ as the fundamental length scale and the units of $C_a$ and $C_R$ as plotted are $2\pi NnR^2 \hbar^2/M$ and $2\pi Nn\hbar^2/M$, respectively.}
\label{cacr}
\end{figure}

In this subsection, we calculate the contact parameters at high temperature by virial expansion~\cite{Ho2004,Liu2009}. The pressure $p$ and the inverse volume (area) $1/v$ can be expanded in terms of powers of fugacity $z=e^{\mu/T}$~\cite{Huang}
\begin{align}
\frac{p}{T} &=\frac{1}{\lambda^2}\sum_l b_l z^l,\\
\frac{1}{v} &=N/V=\frac{1}{\lambda^2} \sum_l l b_l z^l.
\end{align}
here $T$ is temperature, $\lambda=\sqrt{2\pi/T}$ ($k_B=\hbar=1$) is the de Broglie wavelength and $\mu$ is chemical potential.
The grand potential is
 \begin{eqnarray}
\Omega=-pV=-TV\frac{1}{\lambda^2}\sum_l b_l z^l.
\end{eqnarray}
The equation of state can be obtained by eliminated the fugacity $z$
  \begin{eqnarray}
\frac{pv}{T}=\sum^{\infty}_{l=1} a_l(T)(\frac{\lambda^2}{v})^{l-1},
\end{eqnarray}
where $a_l$ is virial coefficient. For example,
\begin{eqnarray}
&&a_1=b_1=1,\notag\\
&&a_2=-b_2,\notag\\
&&a_3=4b_{2}^{2}-2b_3,\notag\\
&&a_4=-20b_{2}^3+18b_2b_3-3b_4.
\end{eqnarray}
For 2D ideal Fermi gas, the inverse area $\frac{1}{v}=N/V=\sum_k\frac{ze^{-\beta k^2 /2}}{1+z e^{-\beta k^2/2}}=\frac{ln(1+z)}{\lambda^2}=\frac{1}{\lambda^2}\sum_n (-1)^{n-1}\frac{z^n}{n}$, so $b^{0}_{l}=(-1)^{l-1}/l^2$. Including the interaction effects, the correction to the second virial coefficient is given by
\be
\Delta b_2=4 \left\{\sum_{b }e^{-E_b/T}+\frac{1}{\pi}\int_0^{\infty}dk\frac{\partial\delta_1}{\partial k}e^{-k^2/T}\right\}.
\ee
The summation on the right hand side includes the effects of possible bound states while the integral takes into account the scattering fermions. The phase shift $\delta_1$ is given before and explicitly,
\begin{equation}
\frac{\partial \delta_1(k)}{\partial k}=-\frac{2\left(\frac{1}{ak^3}+\frac{1}{\pi k}\right)}{1+\left(-\frac{1}{ak^2}+\frac{2 \ln(R k)}{\pi}\right)^2}.
\end{equation}
The contacts can be obtained from thermodynamics potential $\Omega$
\begin{align}
\frac{\partial \Omega}{\partial(a^{-1})} &=-\frac{\pi}{2}C_{a}=-TV\frac{z^2}{\lambda^2}\frac{ \partial \Delta b_2 }{\partial a^{-1}},\\
 \frac{\partial \Omega}{\partial(lnR)}& =C_{R} =-TV\frac{z^2}{\lambda^2}\frac{ \partial \Delta b_2 }{\partial ln R},
\end{align}
where $z$ can be obtained approximately from the equation of state for a classical ideal gas, e.g,  $z=e^{n\lambda^2}-1\simeq n\lambda^2$. So one finally obtains
\begin{align}
C_{a} &=\frac{2}{\pi}TVn^2\lambda^2\frac{ \partial \Delta b_2 }{\partial a^{-1}}=\frac{2}{\pi}2\pi Nn\frac{ \partial \Delta b_2 }{\partial a^{-1}},\label{caHT}\\
C_{R} &=-TVn^2\lambda^2\frac{ \partial \Delta b_2 }{\partial ln R}=-2\pi N n\frac{ \partial \Delta b_2 }{\partial ln R}\label{crHT},
\end{align}
here $N$ is particle number and $n$ is particle density. Fig.\ref{cacr} shows how the second virial co-efficient $\Delta b_2$ and the contacts $C_{a}$ and $C_{R}$ changes as a function of interaction parameters $1/a$ {at fixed  $R$}. Because $C_{a}\geq0$, $\Delta b_2$ is monotonically increasing with $a^{-1}$. Generally speaking, close to resonance $1/a\sim 0$, the interaction effects $\Delta b_2$ diminishes with the increasing of temperature. Because the contribution of bound states to the contact $C_{R}$ is negative, with increasing attractive interaction ($1/a\to 0_+$), the contact $C_R$ decreases and eventually becomes negative.

\section{Breathing mode and contacts}
\label{bm}
It is known that for scaling invariant interactions in two dimensions: $U(\lambda \rho)=U(\rho)/\lambda^2$, where $\lambda$ is an arbitrary scaling factor, the system has $SO(2,1)$ symmetry and the frequency of breathing mode in harmonic trap is exactly twice the trap frequency $2\omega$~\cite{Pitaevskii}. This applies, for example, for the cases of delta contact $\delta^2(\vec r_1-\vec r_2)$ or inverse squared potential $1/|\vec r_1-\vec r_2|^2$.  However, the true inter-atomic interactions break the scaling invariance, and this leads to the frequency shift of breathing modes. In two-component Fermi gas with $s$-wave interactions, one finds that this shift is related to the contact of the system~\cite{Hofmann, Gao}, and in particular, at unitarity, the scaling invariance is regained~\cite{Ed}. However, quite differently for a single component Fermi gas with $p$-wave interaction in two dimensions, the scaling invariance is also broken even at resonance when $a\to \infty$ due to the existence of the contact $C_R$. In the following, we calculate the breathing mode frequency shift and relate it to the two $p$-wave contacts $C_{a}$ and $C_{R}$.

The Hamiltonian in trapped system is given by $H=H_0+V_{\rm trap}+H_{\rm int}$, where $H_0=\frac{1}{2}\sum_i p_{i}^{2}$ is the kinetic energy (recall $M=\hbar=1$). We consider isotropic harmonic trap in two-dimension
\begin{equation}
V_{\rm trap}=\frac{\omega^2}{2}\sum_i r_{i}^{2}
\end{equation}
where $\omega$ is the trapping frequency. The interaction between the atoms $H_{\rm int}$ is of the standard form $H_{\rm int}=\sum_{i<j} U(\vec{r}_i-\vec{r}_j)$. Introducing excited operator of breathing modes
\begin{equation}
 O\equiv\sum_i r^{2}_{i}=\frac{2}{\omega^2}V_{\rm trap}.
\end{equation}
The equation of motion of operator $O$
\begin{align}
i\frac{\partial O(t)}{\partial t}=[O,H]=\sum_i[\vec r^{2}_{i},\vec p^{2}_{i}/2]=2iD.
\end{align}
Here $D\equiv\sum_i [\vec{r}_{i}\cdot\vec{p} _{i}+\vec{p}_{i}\cdot\vec{r} _{i}]/2$ is the dilation operator.
The equation of motion of $D$ is given by
 \begin{align}
i\frac{\partial D}{\partial t}=[D,H]=[D,H_0+V_{\rm trap}+H_{\rm int}].
\end{align}
The various commutators can be evaluated explicitly. In particular, we have
\begin{align}
[D,H_0] &=2i H_0,\\
 [D,V_{\rm trap}] &=-2i V_{\rm trap}
\end{align}
and
\begin{align}
[D,H_{int}]=-i\sum_{i<j}[ \vec{r}_{ij}\cdot\frac{\partial }{\partial \vec{r}_{ij}}]U(\vec{r}_i-\vec{r}_j),
\end{align}
here $\vec{r}_{ij}=\vec{r}_i-\vec{r}_j$. As a result, one finds that
\begin{align}
&\frac{\partial^2 O(t)}{\partial t^2}=2\frac{\partial D}{\partial t}\\
&=4 H_0-4 V_{trap}-2\sum_{i<j}\left[ \vec{r}_{ij}\cdot\frac{\partial }{\partial \vec{r}_{ij}}\right]U(\vec{r}_i-\vec{r}_j)
\end{align}
Using the fact that $\hat{\rho}_2(\vec{r},t)\equiv2\sum_{i<j}\delta^2 (\vec{r}(t)-\vec{r}_{ij})$, we can write down the equation of motion for the average value of $O$ as
\begin{equation}
\frac{\partial^2 \langle O(t)\rangle}{\partial t^2}\!\!=\!\!4 \langle H\rangle-4\omega^2\langle O\rangle\!-\!\!\int \!\!d^2\vec r \left[2U(r)\!\!+\!\!\vec{r}\cdot\frac{\partial U(r)}{\partial \vec{r}}\right]\rho_2(r,t).\label{oeom}
\end{equation}
In the case of scaling invariant potential in two-dimensions, $2U(r)+\vec{r}\cdot\frac{\partial U(r)}{\partial \vec{r}}\equiv0$, the interaction effects on the breathing mode frequency shift vanish exactly. In the appendix, we show that the interaction correction can be written in terms of $p$-wave contacts
\begin{equation}
\int d^2\vec r \left[2U(r)+\vec{r}\cdot\frac{\partial U(r)}{\partial \vec{r}}\right]\left\langle\hat{\rho}_2(r,t)\right\rangle\!\!=\!\!-\frac{2\pi C_{a}(t)}{a}-2C_{R}(t)\label{sa}.
\end{equation}
As a result, the equation of motion for the breathing mode operator $O$ is given by, when taking the expectation value on the many-body state
\begin{equation}
\frac{\partial^2 \langle O(t)\rangle}{\partial t^2}=4 E-4\omega^2 \left\langle O(t)\right\rangle+\frac{2\pi C_{a}(t)}{a}+2C_{R}(t).
\end{equation}
The time-dependences of the contacts gives rises to the correction of the breathing mode frequency.
We note that the average energy   $E\equiv\left\langle H\right\rangle$ does not dependent on time $t$ because of $[H,H]=0$. In a stationary state, the virial theorem is recovered
\begin{equation}
2V_{\rm trap}= E+\frac{\pi}{2a}C_{a}+\frac{C_{R}}{2}.
\end{equation}
When the $p$-wave contacts are zero, the scaling invariance is restored and the frequency of breathing mode is exactly $2\omega$. Near the resonance $1/a\to 0$, although $C_{a}$ term vanishes, $C_{R}$ is still finite and this breaks the scaling invariance even at resonance.

\begin{figure}[H]
\begin{center}
\includegraphics[width=\columnwidth]{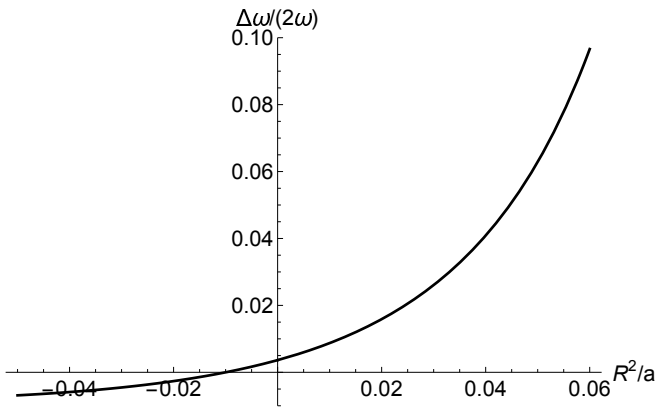}
\end{center}
\caption{(Color online.) The frequency shift $\Delta\omega/(2\omega)$ of breathing modes as a function of $R^2/a$ near the resonance.}
\label{breathingmode}
\end{figure}

In the following, we will investigate the breathing mode frequency shift in the high temperature limit. We write the density distribution in trap during the breathing motion of the cloud in the following scaling form
\begin{equation}
n(\vec r,t)=\frac{1}{\gamma^2(t)}n_0\left(\frac{\vec{r}}{\gamma(t)}\right)
\end{equation}
where $\gamma(t)=1+\Delta \gamma(t)$, $\Delta\gamma(t)\ll1$ for small oscillations. Here $n_0(\vec{r})$ is density distribution of equilibrium  and particle number $N=\int d^2\vec r n_0(\vec r)$. On the other hand, the expectation value of the breathing mode  operator can be written as
\begin{align}
\langle O(t)\rangle &=\int d^2\vec r r^2 n(\vec{r},t)=\gamma^2(t)\int d^2\vec r r^2n_0(\vec r) \\
&\simeq[1+2\Delta\gamma]\langle O\rangle,
\end{align}
here $\langle O \rangle\equiv\int d^2\vec r r^2n_0(\vec r)$ is the average value of $O$ in equilibrium.
Similarly, from high temperature virial expansion, eqns.(\ref{caHT},\ref{crHT}) and using local density approximation, we get
\begin{align}
C_{a}(t) &= \frac{\langle C_{a}\rangle}{\gamma^2},\\
C_{R}(t) &=\frac{\langle C_{R}\rangle}{\gamma^2}.
\end{align}
Here $\langle C_{a}\rangle\equiv \frac{2}{\pi}T\lambda^2\frac{ \partial \Delta b_2 }{\partial a^{-1}}\int d^2\vec r n^{2}_{0}(\vec r)$ and $\langle C_{R}\rangle\equiv -T\lambda^2\frac{\partial \Delta b_2 }{\partial \ln R}\int d^2\vec r n^{2}_{0}(\vec r)$ are contact values in equilibrium.
\begin{equation}
\frac{\partial ^2 \Delta\gamma(t)}{\partial t^2}+\left[4\omega^2+\frac{{2\pi \langle C_{a}\rangle}/{a}+2\langle C_{R}\rangle}{\langle O\rangle}\right] \Delta\gamma(t)=0.
\end{equation}
One thus finds that for small breathing motion of the cloud, the frequency is given by
\begin{equation}
\omega_{\rm b}=\sqrt{4\omega^2+\frac{{2\pi \langle C_{a}\rangle}/{a}+2\langle C_{R}\rangle}{\langle O\rangle}}.
\end{equation}
Assuming that the shift is small, one can expand $\omega_{\rm b}$ and finds
\begin{equation}
\Delta\omega\equiv \omega_{\rm b}-2\omega=\frac{1}{2\omega\langle O\rangle}\left[\frac{\pi \langle C_{a}\rangle}{a}+\langle C_{R}\rangle\right].
\end{equation}
One can evaluate the frequency shift at high temperature where density distribution can be approximated by the classic Boltzmann distribution $n_0(\vec r)=A\exp{(-{\omega^2 \vec r^2}/{2T}})$ where $A={N\omega^2}/({2\pi T})$ and $N$ is particle number.
So the frequency shift becomes
\begin{equation}
\Delta\omega =\frac{N\omega^3}{8T^2}\left[\frac{2}{a}\frac{ \partial \Delta b_2 }{\partial a^{-1}}-\frac{ \partial \Delta b_2 }{\partial \ln R}\right].
\end{equation}
Near the resonance $1/a \rightarrow 0$, the frequency shift only arises from the contact of effective range $C_{R}$. Note that unlike the $p$-wave contacts for three-dimensions, $C_{R}$ does not vanish at resonance. {Here we take $T=2E_F$ with Fermi energy related to radial trap frequency $E_F=\hbar \omega \sqrt{2N}$ in two dimension and}  the breathing mode frequency shift $\Delta\omega$ at high temperature $T=0.1/R^2$ is given in Fig.\ref{breathingmode}. Note that it is finite at resonance when {$1/a=0$}, and can be negative in the {BCS} side of the resonance {when $1/a<0$}. In the BEC side of resonance $1/a>0$, the frequency shift is always positive and becomes larger and larger as the attractive interaction becomes strong ($a\rightarrow 0+$).

\section{Summary}
\label{summary}
In this paper, we have extended the concepts of $p$-wave contacts to two dimensions and derived the related universal relations. In particular, we have shown how the $p$-wave contacts change the breathing mode frequency from the scaling invariant result. Specific results are obtained in the high temperature limit where the 2D system is more stable. It would be interesting if further experiments in two dimension can be conducted and in the appendix, we consider realistic trap parameters and show how the effective two dimensional scattering parameters depends on the magnetic field.

\section*{Acknowledgements.}
We would like to thank  Zhenhua Yu, Joseph H. Thywissen for useful discussions. This research is supported by Hong Kong Research Grants Council (General Research Fund, HKU 17306414, 17318316 and Collaborative Research Fund, HKUST3/CRF/13G) and the Croucher Foundation under the Croucher Innovation Award.


\appendix
\section{Derivation of Eq.(\ref{sa})}
According to the definition of the $p$-wave contacts, the two-body density matrix $\rho_2(\vec r_1,\vec r_2; \vec r_2,\vec r_1) (\equiv\rho_2(\vec{r}_1-\vec r_2))$ in the short-range  can be written as,
\begin{equation}
\rho_2(r)=\frac{C_{a}}{\pi}u^2_{0}(r)+\frac{2C_{R}}{\pi}u_0(r)u_1(r),
\end{equation}
where $r=|\vec r_1-\vec r_2|$. This gives the possibility of finding two two-particle apart distance $r$.
So eqn.(\ref{oeom}) can be written as
\begin{equation}
\frac{\partial^2 \langle O(t)\rangle}{\partial t^2}= 4 \langle H\rangle-4\omega^2\langle O(t)\rangle-(A+B),
\end{equation}
where
\begin{align}
A &=2C_{a}\int d r r\left[2U(r)+r\frac{\partial U(r)}{\partial r}\right]u^{2}_{0}(r)\\
B &=2C_{R} \int dr r \left[2U(r)+r\frac{\partial U(r)}{\partial r}\right]2u_0(r)u_1(r)
\end{align}
Our task below will be to evaluate the above two expressions. One useful fact to notice is that since $U(r)$ is a short-range function, the integrals are effectively cutoff at $r_c$ because $U(r)=0$ when $r\geq r_c$. As a result, we can write $A$ as and integrating from $0$ to $r_c$, we find
\begin{align}
A &=2C_{a}\int d r r\left[2U(r)+r\frac{\partial U(r)}{\partial r}\right]u^{2}_{0}(r)\notag\\
&=-4C_{a} \int dr r^2 U(r)u_0(r)\frac{\partial u_0(r)}{\partial r}.
\end{align}
In the above derivation, we have integrated second term by part and used the fact that $U(r)r^2u_{0}^{2}(r)|_{0}^{r_c}=0$. Using differential equation for $u_0(r)$
\begin{equation}
\left[\frac{1}{r}\frac{\partial}{\partial r}\left(r\frac{\partial}{\partial r}\right)-\frac{1}{r^2}\right]u_0(r)=U(r)u_0(r),
\end{equation}
and multiply both side by $r^2 u_0'(r)$ and integrate from $0$ to $r_c$, we find, using $u_0(r=0)=0$ and $r\frac{\partial u_0(r)}{\partial r}|_{r=0}=0$, and eqn.(\ref{u0})
\begin{align}
A &=-2C_{a}\left[\left(r\frac{\partial u_0(r)}{\partial r}\right)^2-u^{2}_{0}(r)\right]_{0}^{r_c}\notag\\
&=-2C_{a}\left[\left(\frac{1}{r_c}+\frac{\pi r_c}{4a}\right)^2-\left(-\frac{1}{r_c}+\frac{\pi r_c}{4a}\right)^2\right]\notag\\
&=-2\pi\frac{C_{a}}{a}.
\end{align}

The calculation of $B$ needs more efforts. Let us first consider the following integral
\begin{align}
2C_{R} \int dr r \left[2U(r)+r\frac{\partial U(r)}{\partial r}\right]R^2_{k}(r)
\end{align}
where $R_{k}(r)=u_0(r)+k^2u_1(r)$ is the radial wave function expanded to second order in $k$. $B$ is simply given by the coefficient of $k^2$ in the above integral. $R_k(r)$ obeys the following differential equation
 \begin{equation}
 \left[\frac{1}{r}\frac{\partial}{\partial r}\left(r\frac{\partial}{\partial r}\right)-\frac{1}{r^2}\right]R_k(r)+k^2R_k(r)=U(r)R_k(r).
\end{equation}
Multiply both side by $r^2R'_k(r)$ and integrate it by part, one finds
\begin{align}\nonumber
&\int_0^{r_c}U(r)r^2R'_k(r)R_k(r)dr =k^2\int_0^{r_c}r^2R'_k(r)R_k(r)dr\\
&+\frac{1}{2}\left[\left(r\frac{\partial R_k(r)}{\partial r}\right)^2-R^{2}_{k}(r)\right]_{0}^{r_c}.\label{Rpartial}
\end{align}
Similar for $A$, we have
\begin{align}
&\int dr r \left[2U(r)+r\frac{\partial U(r)}{\partial r}\right]R^2_{k}(r)\notag\\
=&-2\int dr r^2 U(r)R_k(r)\frac{\partial R_k(r)}{\partial r}\notag\\
=&-2k^2\int_0^{r_c}r^2R'_k(r)R_k(r)dr\notag\\
&-\left[\left(r\frac{\partial R_k(r)}{\partial r}\right)^2-R^{2}_{k}(r)\right]_{0}^{r_c}
\end{align}
The coefficient of $k^2$ in the above expression determines $B$ (up to a factor $2C_R$). Now, since $[(r\frac{\partial R_k(r)}{\partial r})^2/2-(R^{2}_{k}(r))^2/2]|_{r=0}=0$  and $R_k(r_c>r_0)=u_0(r_c)+k^2u_1(r_c)$, the coefficients of $k^2$ is given
 \begin{eqnarray}
-4C_{R}\!\!\left[\frac{1}{2}\!-\!\frac{\pi^2r^{4}_{c}}{64a^2}\!\!+\!\!\ln\frac{r_c}{\tilde{R}}\!\!+\!\!\int_{0}^{r_c}dr r^2u_{0}(r)u'_{0}(r)\right].
\end{eqnarray}
The later integral $\int_{0}^{r_c}dr r^2u_{0}(r)u'_{0}(r) $ can be calculated from the Bethe's integral formula. We first transform the limit of integration $r_c\rightarrow \lambda r_c$,
\begin{align}
\int^{\lambda r_c}_{0} ru^{2}_{0}(r)dr=\ln\frac{e^{\gamma_E}\lambda r_c/2}{R}-\frac{\pi \lambda^2 r^{2}_{c}}{4a}+\frac{ \pi^2\lambda^4 r^{4}_{c}}{64a^{2}}.
\end{align}
Then we can change the integral variable $r=\lambda x$
\begin{align}
\int^{ r_c}_{0} xu^{2}_{0}(\lambda x)dx=\ln\frac{e^{\gamma_E}\lambda r_c/2}{R}/\lambda^2-\frac{\pi  r^{2}_{c}}{4a}+\frac{ \pi^2\lambda^2 r^{4}_{c}}{64a^{2}}.
\end{align}
Taking the derivative with respect to $\lambda$ and then setting $\lambda=1$, we get
\begin{align}
\int^{ r_c}_{0} x^2 u_{0}( x)u'_{0}( x)dx=-\ln\frac{e^{\gamma_E} r_c/2}{R}+\frac{1}{2}+\frac{ \pi^2 r^{4}_{c}}{64a^{2}}.
\end{align}
Finally we obtain
 \begin{equation}
A+B=-\frac{2\pi}{a}C_{a}-2C_{R}.
\end{equation}
and
 \begin{equation}
\frac{\partial^2 \langle O(t)\rangle}{\partial t^2}= 4 E-4\omega^2\langle O(t)\rangle+\frac{2\pi}{a}C_{a}(t)+2C_{R}(t).
\end{equation}

\section{The relation between the effective scattering parameters ($a$ and $R$) in two-dimension and the three-dimensional $p$-wave scattering parameters}
It is suggested that the 2D strongly interacting $p$-wave Fermi gas might be more stable than its 3D counterpart~\cite{Levinsen2008}. The 2D fermion gas may be produced from 3D fermion gas by using strong harmonic confinement along $\hat{z}$-direction~\cite{Idziaszek,Peng}. Let us then consider two spinless fermions moving in a harmonic potential of the form
\begin{equation}
V_z(x,y,z)=\frac{1}{2}M\omega_{z}^{2}z^2.
\end{equation}
Thus the potential energy of the two particles are $\frac{1}{2}(2M)\omega_{z}^{2}Z^2+\frac{1}{2}\frac{M}{2}z^2$, where $2Z=z_1+z_2$ is twice the center of mass $z$-co-ordinate and the $z=z_1-z_2$ is the relative co-ordinates. Since the center of mass motion is separated in the harmonic trap, one can write the Hamiltonian for the relative motion
\begin{equation}
H_{\rm rel}=\frac{1}{2\mu}(p_x^2+p_y^2+p_z^2)+\frac{1}{2}\mu\omega_{z}^{2}z^2,
\end{equation}
where $\mu=M/2$ is the effective mass. In the following, we use units such that $M=1$ and so $2\mu=M=1$. We introduce $\vec{\rho}\equiv x\hat{x} +y\hat{y}$ as the $x-y$ plane projection of relative coordinate $\vec{r}=\vec{r}_1-\vec{r}_2$ and $\rho=\sqrt{x^2+y^2}$. The free Green's function for the relative motion is given by
\begin{equation}
G^{\pm}(\vec{r},\vec{r}'; E)=\langle \vec{r}_1|\frac{1}{E\pm i\eta-H_{\rm rel}}|\vec{r}_2\rangle.
\end{equation}
In the following, we shall shift the reference point of energy $\epsilon\equiv E-\omega_z/2$. We are interested in the scattering of two fermions in the $xy$-plane, so let us set $z_1=z_2=0$ and introduce a complete set of harmonic oscillator states
\begin{align}
G^{\pm}(\rho)\equiv\frac{1}{V}\sum_{k,n}\frac{e^{i\vec{k}\cdot\vec{\rho}}\phi_n(0)\phi^{*}_{n}(0) }{\epsilon\pm i\eta-(k^{2}_{x}+k^{2}_{y}+n\omega_z)}.
\end{align}
Here $\vec{k}=(k_x,k_y)$ lies in the $xy$-plane. $G^{\pm}(\rho)$ describes the out-going and in-coming waves in two-dimensions. Harmonic oscillator wave functions
\begin{equation}
\phi_{n}(z)=\left[\frac{\alpha}{\sqrt{\pi 2^n n!}}\right]^{1/2}H_{n}(\alpha z),
\end{equation}
where $\alpha=\sqrt{\mu \omega_z/\hbar}=\sqrt{\omega_z/2}$. The scattering wave function of the relative motion can then be written as
\begin{equation}
\psi=\frac{\partial G^{-}(\rho)}{\partial \rho}-S\frac{\partial G^{+}(\rho)}{\partial \rho}.
\end{equation}
The stationary scattering wave function of the relative motion is a superposition of the incoming and the outgoing waves, the coefficient of which is the diagonalized S-matrix element $S=\exp{(2i\delta_1)}$ in terms of the scattering phase shift $\delta_1$. In the following, we consider low-energy scattering when $0<\epsilon<\omega_z$. To proceed further, we split the summation over $n$ into two parts, one with $n=0$ and another with $n>0$,  $G^{\pm}(\rho)\equiv G^{\pm}_{1}(\rho)+G^{\pm}_{2}(\rho)$, with
\begin{align}
 G^{\pm}_{1}(\rho) &\equiv\frac{1}{V}\sum_{k}\frac{e^{i\vec{k}\cdot \vec{\rho}}\phi_0(0)\phi^{*}_{0}(0) }{\epsilon\pm i\eta-(k^{2}_{x}+k^{2}_{y})},\\
 G^{\pm}_{2}(\rho) &\equiv \frac{1}{V}\sum_{k,n>0}\frac{e^{i\vec{k}\cdot \vec{\rho}}\phi_n(0)\phi^{*}_{n}(0) }{\epsilon\pm i\eta-(k^{2}_{x}+k^{2}_{y}+n\omega_z)}.
 \end{align}
Using the form of $\phi_n(0)$ and carry out integrations, one finds,
\begin{align}
 G^{\pm}_{1}(\rho) &=\sqrt{\frac{\omega_z}{2\pi}}\frac{\mp i [J_0(\sqrt{\epsilon}\rho)\pm i N_0(\sqrt{\epsilon}\rho)]}{4},\\\nonumber
 G^{\pm}_{2}(\rho) &=-\sqrt{\frac{\omega_z}{2\pi}}\frac{1}{4\pi}\int_{0}^{\infty}\frac{dt}{t}\exp\left[{\frac{-\omega_z\rho^2}{2t}+\frac{\epsilon t}{2\omega_z}}\right]\\
& \times\left[\frac{1}{\sqrt{1-e^{- t}}}-1\right].
 \end{align}
Note that for $G_2^{\pm}$, it is independent of the sign of the infinitesimal imaginary part in the denominator and we shall denote it as $G_2$ simply in the following. Let $\tilde{G}_2=\partial G_2/\partial \rho$,
\begin{align}\nonumber
 \tilde{G}_{2}(\rho) &=\sqrt{\frac{\omega_z}{2\pi}}\frac{\omega_z \rho}{4\pi}\int_{0}^{\infty}\frac{dt}{t^2}\exp\left[{\frac{-\omega_z\rho^2}{2t}+\frac{\epsilon t}{2\omega_z}}\right]\\
& \times\left[\frac{1}{\sqrt{1-e^{- t}}}-1\right].
 \end{align}
Then the wave function can be written as
\begin{align}
\psi &=\sqrt{\frac{\omega_z}{2\pi}}\frac{- i\sqrt{\epsilon} [J_1(\sqrt{\epsilon}\rho)- i N_1(\sqrt{\epsilon}\rho)]}{4}+\tilde{G}_2\notag\\
 &-S\left[\sqrt{\frac{\omega_z}{2\pi}}\frac{ i\sqrt{\epsilon} [J_1(\sqrt{\epsilon}\rho)+ i N_1(\sqrt{\epsilon}\rho)]}{4}+\tilde{G}_2\right],\notag\\
 &\propto J_1(\sqrt{\epsilon}\rho)-\tan\delta_1N_{1}(\sqrt{\epsilon}\rho)+\frac{\tan\delta_{1}}{A}\tilde{G}_2. \label{wave1}
\end{align}
Here $A\equiv\sqrt{\frac{\omega_z}{2\pi}}\frac{\sqrt{\epsilon}}{4}$. From now, we set $k\equiv\sqrt{\epsilon}$.
When $\rho^2\omega_z\ll1$, the wave function looks like 3D p-wave function.
\begin{align}
\psi &\propto j_1(p\rho)-\tan(\delta^{3D})n_1(p\rho),\notag\\
&\propto \cot(\delta^{3D})j_1(p\rho)-n_1(p\rho),\notag\\
&\propto \left[-\frac{1}{3V^{3D}}-\frac{p^2}{3R^{3D}}\right](\rho+\cdots)+\left[\frac{1}{\rho^2}+\cdots\right].\label{wave2}
\end{align}
Here $V^{3D}$ and $R^{3D}$ are $p$-wave scattering volume and effective range in three-dimension, respectively~\cite{Yu2015}. The energy $E=p^2=k^2+\omega_z/2$. In the above equation, the effective expansion for 3D p-wave phase shift $p^3\cot(\delta^{3D})=-1/V^{3D}-p^2/R^{3D}$ has been used.  $j_1$ and $n_1$ are spherical Bessel functions. At small $\rho$, we need to calculate the coefficients of $\rho$ and $\rho^{-2}$ from $J_1$, $N_1$ and $\tilde{G}_2$, and then compare the above two formulas, to get the effective 2D p-wave interaction parameters in terms of 3D p-wave parameters $V^{3D}$ and $R^{3D}$.
As $\rho\rightarrow0$, the most divergence term is proportional $1/\rho^2$ which comes from
\begin{align}
\tilde{G}_2 &\sim\sqrt{\frac{\omega_z}{2\pi}}\frac{\omega_z\rho}{4\pi}\int_{0}^{\infty}\frac{dt}{t^2}\exp\left[{\frac{-\omega_z\rho^2}{2t}+\frac{\epsilon t}{2\omega_z}}\right]\sqrt{\frac{1}{t}},\notag\\
&\sim \frac{1}{4\pi\rho^2}.
\end{align}
As a result, the coefficient of $1/\rho^2$ is $\frac{\tan\delta_1}{4\pi A}$ in eqn(\ref{wave1}).

Next we calculate the coefficient of $\rho$. There exists $\rho \ln\rho$ and $1/\rho$ terms in the expansion of Bessel's function $N_1(k\rho)$ near origin. However there no such terms ($\rho \ln\rho$ and $1/\rho$)  in 3D scattering wave function ($j_1$ and $n_1$). In fact, it can be shown that these singular term from $N_1(kr)$ are exactly canceled by that from $\tilde{G}_2$. Here we need isolate these singular terms by a formula
\begin{align}
\Delta \tilde{G}_2 &=\tilde{G}_2-g_2\notag\\
  &=\sqrt{\frac{\omega_z}{2\pi}}\frac{\omega_z \rho}{4\pi}\int_{0}^{\infty} \frac{dt}{t^2}\exp\left[{\frac{-\omega_z\rho^2}{2t}+\frac{k^2 t}{2\omega_z}}\right]\left[\frac{1}{\sqrt{1-e^{- t}}}\right.\notag\\
  &\left.-1-\frac{e^{-t}}{\sqrt{t}}+e^{-t}-\frac{5\sqrt{t}e^{-t}}{4}+te^{-t}\right].
\end{align}
$g_2$ gives all the mainly singular terms, {\em e.g,} $1/\rho^2,1/\rho,\rho \ln\rho$ occurring at $\tilde{G}_2$ as $\rho\rightarrow 0$. Then $\Delta \tilde{G}_2$ is {not} singular comparing with linear term $\rho$ as $\rho\rightarrow0$, {\em e.g.} $\Delta \tilde{G}_2/\rho \sim O(1)$. When $\rho\rightarrow0$, we have
\begin{equation}
\Delta \tilde{G}_2=\sqrt{\frac{\omega_z}{2\pi}}\frac{\omega_z }{4\pi}\beta(k^2/\omega_z) \rho,
\end{equation}
and
\begin{align}
\beta(k^2/\omega_z)\equiv\int_{0}^{\infty} & \frac{dt}{t^2}\exp\left[\frac{k^2 t}{2\omega_z}\right]\left[\frac{1}{\sqrt{1-e^{- t}}}-1\right.\\\notag
&-\left.\frac{e^{-t}}{\sqrt{t}}+e^{-t}-\frac{5\sqrt{t}e^{-t}}{4}+te^{-t}\right].
\end{align}
Collecting all the linear term of $\rho$ from $J_1, N_1, g_2$ and $\Delta \tilde{G}_2$ and
comparing the coefficients of $\rho$ and $1/\rho^2$ in Eq.(\ref{wave1}) and (\ref{wave2}), we can get the effective $p$-wave interaction parameters $a$ and $R$ in two dimensions in terms of $V^{3D}$ and $R^{3D}$.
\begin{align}
a &=\frac{3\pi V^{3D}}{2\sqrt{2\pi}l_z+\frac{\sqrt{2\pi }V^{3D}}{R^{3D}l_z}+\frac{V^{3D}}{l_{z}^{2}}(6-7\sqrt{\pi}+6\beta(0))},\\
R&=\frac{e^{\frac{3\sqrt{\pi}}{8}-\frac{\gamma}{2}}}{\sqrt{2}}e^{-\frac{\sqrt{2\pi }l_z}{3R^{3D}}}l_z,
\end{align}
here $l_z=\sqrt{\hbar/M\omega_z}$ and $\beta(0)\approx 1.01353$. The coefficient $\gamma$ is given by
\begin{equation}
\gamma \equiv\int_{0}^{\infty} \frac{dt}{t}\left[\frac{1}{\sqrt{1-e^{- t}}}-1-\frac{e^{-t}}{\sqrt{t}}+e^{-t}-\frac{5\sqrt{t}e^{-t}}{4}+te^{-t}\right]
\end{equation}
which is approximately $0.3915$.

\begin{figure}[H]
\begin{center}
\includegraphics[width=\columnwidth]{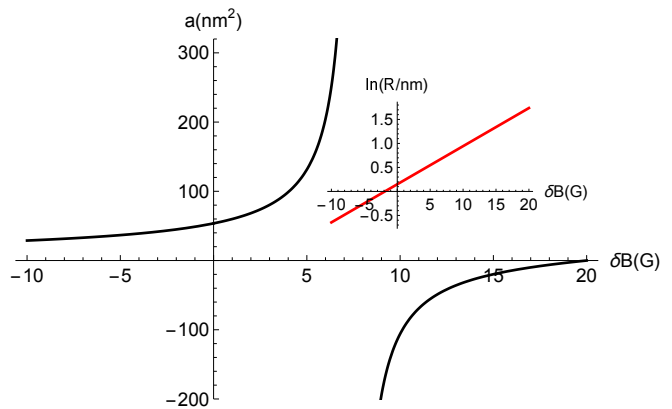}
\end{center}
\caption{(Color online.) This scattering area $a$ as a function of magnetic field $\delta B$ (relative to 3D resonance point). The inset shows the variations of  effective range $R$ with magnetic field $\delta B$.  }
\label{vr}
\end{figure}

In the following, one can choose the  parameters which are experimentally accessible in  $^{40}$K~\cite{Luciuk}. Consider a harmonic trap with frequency of order of $\omega_z=2\pi\times 120$kHz, then use
\begin{equation}
V^{3D}=V_{bg}\left(1-\frac{\Delta_V}{\delta B}\right),
\end{equation}
where the (magnetic) width of the resonance is $\Delta_V=20$G. $V_{bg}=(100a_0)^3$ is the background scattering length. $a_0$ is Bohr radius. For the effective range, we have
\begin{equation}
\frac{1}{R^{3D}}=\frac{1}{R_{bg}}\left(1+\frac{\delta B}{\Delta_R}\right),
\end{equation}
where $\Delta_R=-20G$ and $R_{bg}=50a_0$. The above experimental parameters can realized a quasi-2D Fermi gas which satisfies $R^{3D}\ll l_z\ll 1/k_F$~\cite{Thywissen}. In order to realize true 2D Fermi gas where $R^{3D}\sim l_z \ll1/k_F$ one needs a trap frequency of order of $\omega_z\sim10$MHz. In Fig.\ref{vr}, for $\omega_z=10$MHz, we show the variations of effective 2D scattering area $a$ and effective range $R$ when magnetic field $\delta B$ (relative to 3D resonance point) varies. Note that the resonance position is shifted by the external trap potential.

\end{document}